# Spatial beam self-cleaning in bi-tapered multimode fibers

Xiao-Jun Lin[a], Yu-Xin Gao[b], Jin-Gan Long[a], Jia-Wen Wu[a], Xiang-Yue Li[a], Wei-Yi Hong[a], Hu Cui[a], Zhi-Chao Luo[a], Wen-Cheng Xu[a], and Ai-Ping Luo [a, *]

[a] Guangdong Provincial Key Laboratory of Nanophotonic Functional Materials and Devices, South China Normal University, Guangzhou 510006, China

[b] Department of Mechanical and Electrical Engineering, Shandong Polytechnic College, Jining, Shandong 272067, China

* E-mail address: luoaiping@scnu.edu.cn

**Abstract**—We report the spatial beam self-cleaning in bi-tapered conventional multimode fibers (MMFs) with different tapered lengths. Through the introduction of the bi-tapered structure in MMFs, the input beam with poor beam quality from a high-power fiber laser can be converted to a centered, bell-shaped beam in a short length, due to the strengthened nonlinear modes coupling. It is found that the bi-tapered MMF with longer tapered length at the same waist diameter shows better beam self-cleaning effect and larger spectral broadening. The obtained results offer a new method to improve the beam quality of high-power laser at low cost. Besides, it may be interesting for manufacturing bi-tapered MMF-based devices to obtain the quasi-fundamental mode beam in spatiotemporal mode-locked fiber lasers.

**Keywords**—Bi-tapered fiber, multimode fiber, spatial beam self-cleaning, nonlinear optics.

## 1. Introduction

Since the birth of optical communication, its capacity has grown exponentially, especially with the development of fiber and relevant devices. Such great progress has been achieved with the support of multiplexing and modulation technologies such as wavelength division multiplexing [1], time division multiplexing [2], polarization division multiplexing [3], quadrature amplitude modulation [4], and so on. However, the relatively small core diameter and single-core structure of single-mode fiber (SMF) limit the further improvement of transmission capacity. Space division multiplexing technique [5] shows a possible solution to improve the capacity of communication system, including core division multiplexing in multicore fibers [6, 7] and mode division multiplexing in multimode fibers (MMFs) [8, 9]. The structure of the single-core MMF is closer to the SMF, and its drawing, active or passive device fabrication as well as fiber maintenance are more compatible with the traditional SMF. In the process of optical amplification, each mode can effectively share the pump light, and then improve the pump efficiency. Thanks to the above advantages, mode division multiplexing can achieve lower laying cost and energy consumption, which makes MMF a better choice to increase the communication capacity. In addition, MMF also plays an important role in other applications, such as optical imaging [10], high-power laser [11] and mode-locking devices [12].

Although MMF has been manufactured and reported very early, multimode wave propagation and dynamics in MMF have not been well studied until recent years. As an ideal platform, numerous novel physical phenomena have been observed in MMFs, which made significant and rapid progress in nonlinear optics, including multimode optical soliton [13-15], geometric parametric instability [16], ultrabroad dispersive radiation [17, 18], beam self-cleaning [16, 18-39], and so on. As characteristics of the waveguide, modes are intrinsic solutions of Helmholtz equation under boundary conditions, revealing the transverse pattern of light beam. It is well known that when light propagates along MMF, imperfects of fibers such as diameter fluctuation, micro-bending and refractive index fluctuation are mainly responsible for the modes coupling, where transverse modes exchange their energy randomly and excite unpredictable modes which cause the decline of the beam quality, showing "speckle" pattern. A phenomenon called beam Kerr self-cleaning can help to reverse this situation by redistributing the energy of the modes and finally concentrating most of the energy in one low-order mode (LOM). Spatial beam Kerr self-cleaning is of particular interest because it leads to a self-organization of light which confines most of the energy in a quasi-single transverse mode [16, 18-37, 39]. Thus, the "speckle" beam in MMFs could be improved.

Beam self-cleaning occurs not only in the normal dispersion regime [16, 19-36, 38, 39] but also in the anomalous dispersion regime in MMFs [18, 37]. Pulses with different duration from nanosecond to femtosecond have been used to achieve beam Kerr self-cleaning in various MMFs [16, 18-39]. The most common opinion is that the energy of high-order modes (HOMs) flows into the fundamental mode irreversibly through Kerr nonlinearity in fibers, and finally causes the fundamental mode a dominant role among the modes. Obviously, key to induce this effect is the input power level due to the strong dependence on light power of Kerr nonlinearity. Generally speaking, beam Kerr self-cleaning needs high input peak power (up to kW level) of the pulse, especially in short length MMFs. Such strict prerequisite limits the further development of experiments and its practical applications. Therefore, other ways to develop this technology are always under exploring. Methods like gain amplification and novel structure of MMFs are new attempts [18, 23, 24, 32, 34]. In particular, tapered fiber is an ideal test bed for investigating beam Kerr self-cleaning. Compared to fibers with longitudinally uniform core diameter, tapered fibers can bring higher nonlinearity, which may promote the generation of pure fundamental mode. Moreover, their manufacturing methods are already clear and mature so it's easy to prepare tapered fibers in large quantities in the lab. M. A. Eftekhar et al. investigated the accelerated nonlinear intermodal interactions in core decreasing MMFs, where enhanced nonlinear effects led to ultrabroad spectra and beam self-cleaning [18]. A. Niang et al. used a 9.5 m long tapered Yb-doped MMF to achieve beam self-cleaning with and without gain amplification [32]. They found that the combination of the accelerating self-imaging effect and a dissipative landscape led to beam Kerr self-cleaning in both passive and active configurations. However, the core diameters of the tapered MMFs in these experiments are monotonically decreasing or increasing [18, 32, 34] which makes the tapered MMFs difficult to be compatible with traditional fiber devices in practical applications. On the other hand, monotonically decreasing tapered fibers of meters long need complex manufacturing processes.

In this work, bi-tapered conventional MMFs with different tapered length are used to achieve spatial beam self-cleaning. The speckled beam can be converted into a bell-shaped one after it passes through the bi-tapered MMF by beam Kerr self-cleaning effect. Through introducing the tapered structure, nonlinear coupling among the modes is strengthened and the spatial beam self-cleaning is easier to be realized in a relatively shorter length, compared to MMF with the uniform diameter. The bi-tapered MMF with a longer tapered section presents a better beam self-cleaning effect. During this process, it is also accompanied with spectral broadening. The method provides a possibility to improve the beam quality of high-power laser at low cost.

## 2. Experimental setup

Fig. 1 shows the schematic of the experimental setup. Pulse from a high-power laser is coupled into the bi-tapered MMF to observe the beam self-cleaning effect. The initial excitation conditions, such as the input beam size, influence the threshold power of spatial beam self-cleaning [33]. With appropriate initial excitation conditions, the spatial beam self-cleaning could be achieved. Here, we use a collimator and a lens to collimate and focus the beam from the laser on the face of the MMF with bi-tapered structure, adjusting the distance and incident angle carefully to improve the coupling efficiency. At the output of the MMF, a CCD camera (Goldeye, G-033SWIR TEC1) and an optical spectrum analyzer (OSA, Yokogawa, AQ6317C) are employed to record the output beam profile and the spectrum, respectively.

In our experiments, we use standard 62.5/125 μm graded-index (GRIN) MMFs to make the bi-tapered fibers. There are many methods for manufacturing the tapered structure. Considering the actual facilities in our lab, we adopt the flame-heated taper drawing technique. The waist diameter of the bi-tapered MMFs is fixed at about 10 μm. Because we find that the bi-tapered MMFs with waist diameter less than 10 μm could not bear the high-power laser, which always causes damage to the waist section of the bi-tapered MMFs in the experiments. Whereas for the bi-tapered MMFs with a waist diameter larger than 10 μm, the beam self-cleaning is not obvious enough in a short length. Through adjusting the heating time and the drawing force, we can get bi-tapered MMFs with different tapered length. The detailed structure of the bi-tapered MMF is shown in Fig. 2. During the production, the waist section is always ensured to be the center of the entire fiber. For facilitating the coupling and measurement of the light, the total length of each MMF including the bi-tapered section is as long as 1 m.

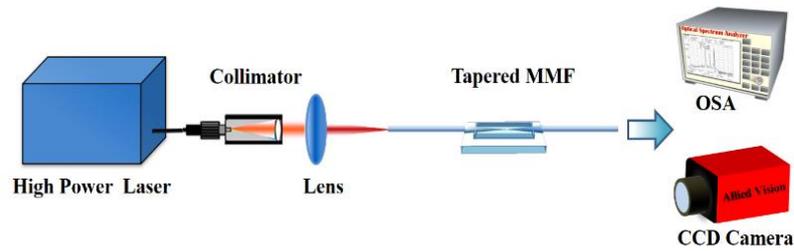

Fig. 1.    Schematic of the experimental setup for observing the spatial beam self-cleaning.

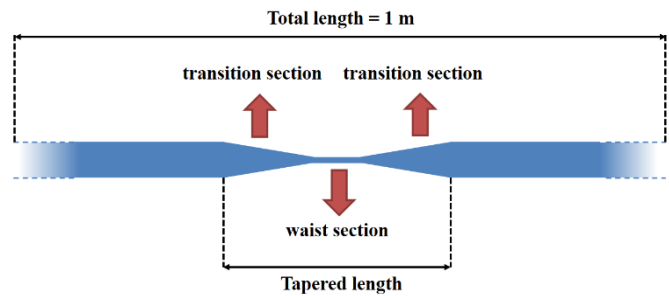

Fig. 2.    Structure of the bi-tapered MMF.

The light source is a home-made high-power Yb-doped fiber laser which launches pulse with 1.26 ps duration at the repetition-rate of 1 MHz. Generally, the input light for beam self-cleaning experiments has a Gaussian-profile [13, 16, 20-37, 39]. Note that the output beam pattern from our laser is not a $LP_{01}$ mode, as shown in Fig. 3. It is because that the light passes several space devices such as focal lens, grating pair and mirrors in the laser amplified system, which causes wave-front distortion and finally leads to the degradation of the beam quality. Nevertheless, it can be converted into a good quality beam through a bi-tapered MMF in a short length.

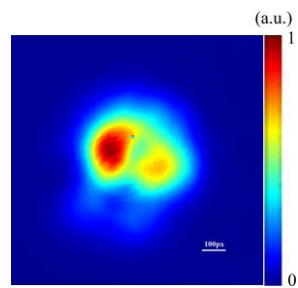

Fig. 3.    Profile of the output beam from the high-power laser (scalebar: 100 pixels).

## 3. Experimental results and discussions

The MMF with a tapered length of 3.6 cm is the first sample for spatial beam self-cleaning experiments. By increasing the input peak power of the pulse, the beam with bad quality from the fiber laser is finally cleaned to a quasi-fundamental-mode beam in the MMFs. As shown in Fig. 4, at relatively low power level, the output beam profiles show disorder or intensity fluctuation as those observed in relevant studies [16, 19-37, 39]. Then, the output beam pattern starts to merge into a single section with the increase of the pulse peak power. The output beam profile from the MMF with the 3.6 cm tapered length concentrates gradually above 119.1 kW. It's clear that the fundamental mode occupies the dominant component during the beam self-cleaning process and the output beam pattern remains stable even the input power continues to increase.

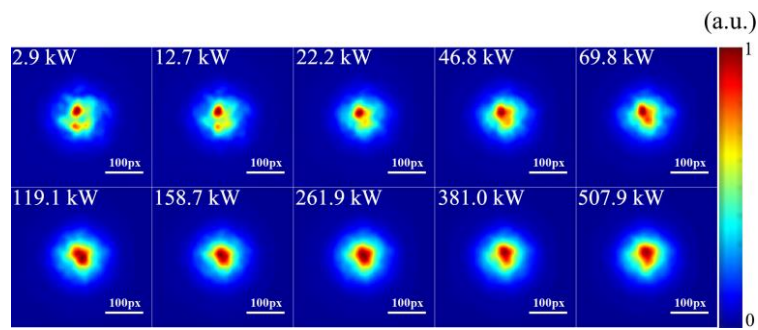

Fig. 4.   Output beam profiles from the MMF with a 3.6 cm tapered length under different input peak powers. (scalebar: 100 pixels)

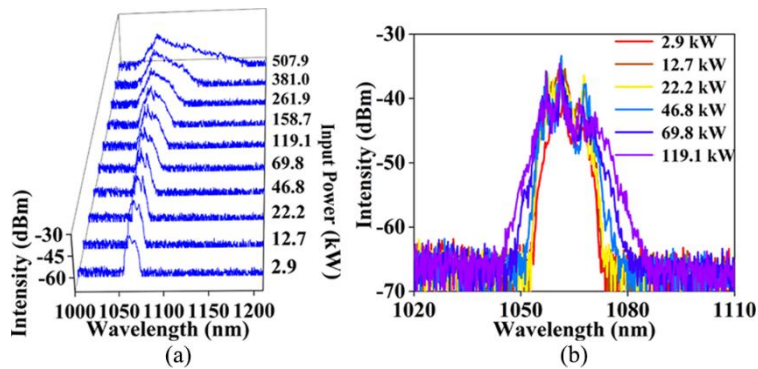

Fig. 5.   (a) Measured output spectra from the MMF with a 3.6 cm tapered length versus input peak powers; (b) the first six spectra.

Note that during the beam self-cleaning, the spectral broadening is often accompanied by the increase of the input power, even generating the supercontinuum [18, 20, 32]. We also measured the corresponding spectral evolution at different input peak powers, as presented in Fig. 5(a). At lower peak power, the spectrum broadens slightly. When the input peak power increases further, the spectrum starts broadening asymmetrically due to the stimulated Raman scattering (SRS) effect, especially, the spectrum broadens to the wavelength near 1190 nm. In order to compare the spectral broadening characteristics more clearly, the first six spectra below 158.7 kW peak power are extracted in Fig. 5(b). As displayed in Fig. 5(b), the spectral broadenings at low peak powers are symmetrical, which is caused by self-phase modulation (SPM).

Noticeably, SPM leads to nonreciprocal coupling for different overlap integrals of the fundamental mode and HOMs [25], which requires sufficiently large power. The beam Kerr self-cleaning starts before the asymmetric broadening of the spectrum. Although the SRS effect takes part in the process, we think that the beam self-cleaning is not a Raman-induced beam cleanup since there are no strong and discrete SRS peaks observed in the spectrum [40], which are typical characteristics of Raman-induced beam cleanup.

For studying the effect of the tapered length with the same waist diameter on the beam self-cleaning, bi-tapered MMFs with different tapered lengths were fabricated for comparison under the same experimental conditions, where the total length of each MMF is kept at 1 m. Figures 6(a) and (b) summarize the output beam profiles from the MMFs with a 1.7 cm and a 5.0 cm long bi-tapered structure, respectively. The MMFs with different tapered lengths achieve beam self-cleaning though the degree of beam self-cleaning differs. For the 5.0 cm long bi-taper, its output beam profile converts into a quasi-fundamental mode above 46.8 kW. Obviously, the longer tapered length introduces better effect of the beam self-cleaning. Moreover, the MMF with a 5.0 cm tapered length shows the best beam self-cleaning result among the three MMFs with bi-tapered structure. Since the beam quality factor $M^2$ cannot be measured accurately due to the device limitation in our lab, the beam intensity distributions in the two orthogonal directions from these three fibers at the input peak power of 507.9 kW are provided and depicted in Fig. 7. It is obvious that the beam transverse distribution from the MMF with a 5.0 cm long bi-taper is closer to a bell-shaped profile, indicating the best beam self-cleaning effect compared with the other two fibers. In addition, the maximum peak power used in our experiment is well below the self-focusing threshold of about 5 MW [41]. Therefore, the self-focusing effect can be neglected.

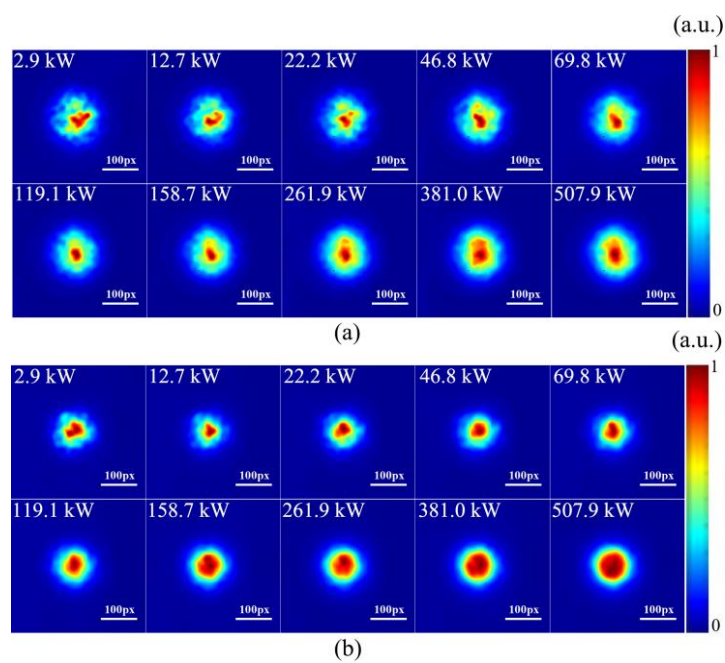

Fig. 6.   Output beam profiles from the MMFs with tapered length of (a) 1.7 cm and (b) 5.0 cm under different input peak powers (scalebar: 100pixels).

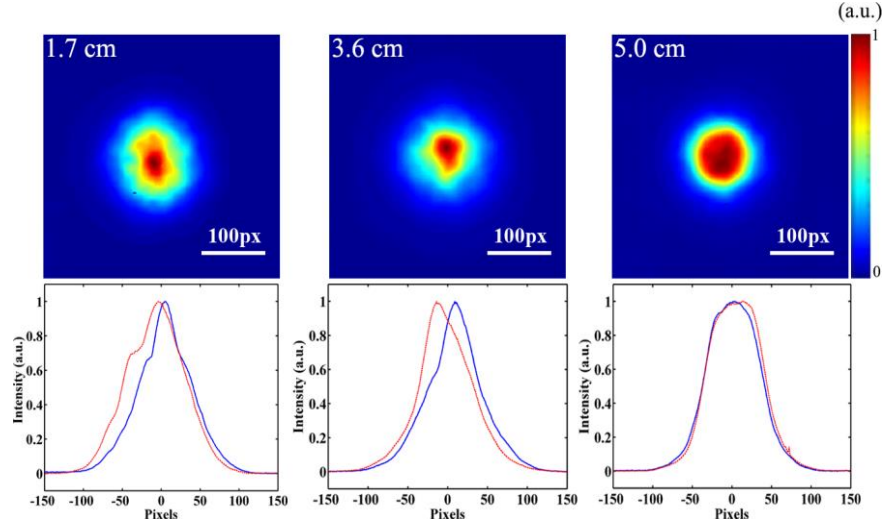

Fig. 7. The beam profiles and the corresponding intensity profiles in the two orthogonal directions (red line: vertical; blue line: horizontal) from the MMFs with different bi-tapered lengths at the input peak power of 507.9 kW (scalebar: 100 pixels).

Furthermore, we recorded the corresponding output spectra in the process of the beam self-cleaning in these two cases. It is found that the broadening process of the spectra is similar to that in Fig. 5. The difference is that the degree of broadening is different at the maximum input peak power, as presented in Fig. 8. The MMF with a 5.0 cm tapered length has the maximum spectral broadening and shows the greatest asymmetry, arising from four-wave mixing and SRS effects working together to generate long-wavelength components. Note that a peak near 1120 nm appears in the spectrum of the MMF with a 5.0 cm long bi-taper, which indicates a more obvious SRS effect compared with the shorter tapered length MMFs. In addition to the output beam profiles and the spectra, the pulses were also recorded. No significant difference is observed. Therefore, the pulses were not considered in this work.

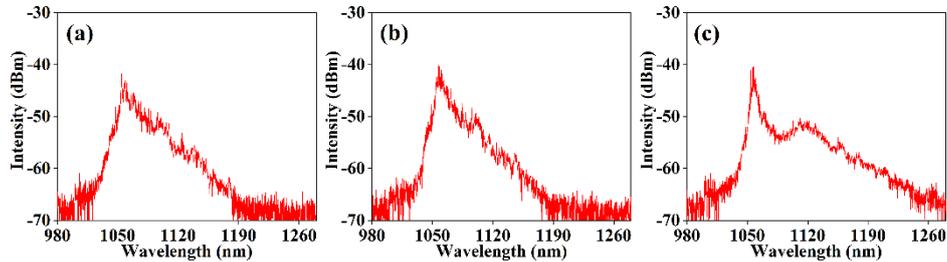

Fig. 8. Measured output spectra from MMFs with tapered length of (a) 1.7 cm, (b) 3.6 cm, and (c) 5.0 cm at the input peak power of 507.9 kW.

To further verify the beam self-cleaning effect of the bi-tapered fiber structure, pulses were launched into a uniform MMF with a total length of 1 m in the same experimental setup. As presented in Fig. 9(a), at low input peak power, a speckled beam is recorded. We did not observe the convincing Kerr-induced or SRS-induced beam self-cleaning even if the input peak power had reached the maximum, as shown in Fig. 9(b). Although there is a tendency for wave condensation in the unstretched MMF but the input peak power is not high enough to reach the threshold power. If the peak power continues to rise (limited by the maximum power of our laser), the output beam from the unstretched MMF may be cleaned to a quasi-fundamental mode [16, 19-22, 25-31, 33]. It further proves that the bi-tapered structure in the MMFs indeed lowers the threshold power of the beam self-cleaning.

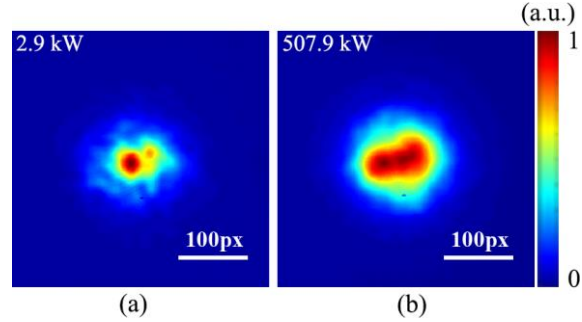

Fig. 9. Output beam profiles from a 1 m long uniform MMF at the input peak power of (a) 2.9kW and (b) 507.9kW (scalebar: 100 pixels).

The spatial beam self-cleaning behavior in MMFs have been explained in many articles [21, 25, 28, 30, 32, 33, 39]. One of the opinions is that Kerr beam self-cleaning originates from nonlinear nonreciprocity [25], indicating irreversibility of energy flowing into the fundamental mode from HOMs above a threshold power. As an important role in these nonlinear behavior, self-imaging effect of GRIN MMF reproduces injected light field periodically and causes the refractive index to modulate along the longitudinal direction due to Kerr effect [21]. This dynamic long-period grating structure in GRIN MMFs favors of quasi phase-matching condition and eventually leads to nonlinear coupling between HOMs and the fundamental mode or LOMs. Finally, the fundamental mode or LOMs occupy the main components of the beam, which means a relatively clean output beam profile. Pedestal (residual) exists even though most of the energy is concentrated in the fundamental mode in the spatial beam self-cleaning process. In addition to the length of the fiber and the input power, the initial excitation condition also influences the cleaning process [28, 29]. For the bi-tapered structure in our experiment, the relatively small core diameter shows stronger nonlinear effect for enhanced light density compared with the unstretched MMF. Therefore, the beam self-cleaning is promoted by the entire taper including the waist section and the transition sections, which could not be seen as a cascading process with monotonically decreasing and increasing tapers [18, 32, 34]. Therefore, the intermodal interaction gets enhanced in tapered region and favors the spatial beam self-cleaning.

For explaining the distinct results in this work, other effects should also be considered. As mentioned above, the spectra from MMFs with bi-tapered structure indicate the presence of the SRS effect. Energy depletion caused by SRS effect hinders the beam Kerr self-cleaning of the pump wave, which will degrade the output beam quality. Actually, the output beam profiles from the MMFs with the bi-tapered structures are quasi-fundamental mode, the experimental results prove that the beam self-cleaning process is mainly induced by Kerr effect. Besides, most of the spectral intensity concentrates in the pump rather than in the Stokes component. In addition, we also considered the influence of the modes number supported in the tapered region. Even though the core size of the waist section is small, it still supports multiple modes propagation by calculating the normalized frequency of the tapered MMF in the waist section [42]. In fact, mode filtering effect in the tapered section is not the reason for the beam self-cleaning phenomena. Otherwise, the output beam profiles would not be much speckled at low input power. Note that the beam self-cleaning may be achieved by nonlinear losses of HOMs such as multi-photon absorption [43]. However, in our experiments, we did not find an obvious drop of the transmitted power through the fiber as in Ref. [43] when the spatial beam self-cleaning occurred. Thus, nonlinear loss can be ignored here.

Different from the longitudinal decreasing or increasing tapered MMFs [18, 32, 34], spatial beam self-cleaning is achieved in shorter conventional MMFs by introducing the bi-tapered structure. The work may offer a possible method to realize beam self-cleaning for MMF laser or other high-power laser through various tapered MMF-based devices. Besides, the tapered MMF may also be hopeful to act as a saturable absorber for MMF lasers.

## 4. Conclusions

In conclusion, we use bi-tapered conventional MMFs to convert a laser beam with poor quality into a quasi-fundamental mode in a short length. The beam self-cleaning process is accompanied by the spectral broadening. The obtained results provide a low-cost method to achieve spatial beam self-cleaning, which may help to improve the high-power laser beam quality and promote the development of other devices based on MMFs.

## Acknowledgement

This work was supported by the National Natural Science Foundation of China under Grants No. 61875058, 92050101, 11874018, 11974006.

## References

[1] L. Mollenauer, A. Grant, X. Liu, X. Wei, C. Xie, I. Kang, Experimental test of dense wavelength-division multiplexing using novel, periodic-group-delay-complemented

dispersion compensation and dispersion-managed solitons, Opt. Lett. 28 (2003) 2043-2045, https://doi.org/10.1364/OL.28.002043.

[2] H. G. Weber, S. Ferber, M. Kroh, C. Schmidt-Langhorst, R. Ludwig, V. Marembert, C. Boerner, F. Futami, S. Watanabe, C. Schubert, Single channel 1.28 Tbit/s and 2.56 Tbit/s DQPSK transmission, Electron. Lett. 42 (2006) 178-179, https://doi.org/10.1049/el:20063873.

[3] X. Liu, F. Buchali, R. Tkach, Improving the nonlinear tolerance of polarization-division-multiplexed CO-OFDM in long-haul fiber transmission, J. Lightwave Technol. 27 (2009) 3632-3640, https://doi.org/10.1109/JLT.2009.2022767.

[4] F. Li, J. Yu, Y. Fang, Z. Dong, X. Li, L. Chen, Demonstration of DFT-spread 256QAM-OFDM signal transmission with cost-effective directly modulated laser, Opt. Express 22 (2014) 8742-8748, https://doi.org/10.1364/OE.22.008742.

[5] D. J. Richardson, J. Fini, L. Nelson, Space division multiplexing in optical fibres, Nat. Photonics 7 (2013) 354-362, https://doi.org/10.1038/nphoton.2013.94.

[6] J. Sakaguchi, B. J. Puttnam, W. Klaus, Y. Awaji, N. Wada, A. Kanno, T. Kawanishi, K. Imamura, H. Inaba, K. Mukasa, R. Sugizaki, T. Kobayashi, M. Watanabe, 305 Tb/s space division multiplexed transmission using homogeneous 19-core fiber, J. Lightwave Technol. 31 (2013) 554-562, https://doi.org/10.1109/JLT.2012.2217373.

[7] R. G. H. Uden, R. Correa, E. Lopez, F. M. Huijskens, C. Xia, G. Li, A. Schulzgen, H. Waardt, A. M. J. Koonen, C. Okonkwo, Ultra-high-density spatial division multiplexing with a few-mode multicore fibre, Nat. Photonics 8 (2014) 865-870, https://doi.org/10.1038/nphoton.2014.243.

[8] S. Randel, R. Ryf, A. Sierra, P. Winzer, A. Gnauck, C. Bolle, R.-J. Essiambre, D. Peckham, A. McCurdy, J. R. Lingle, 6×56-Gb/s mode-division multiplexed transmission over 33-km few-mode fiber enabled by 6×6 MIMO equalization, Opt. Express 19 (2011) 16697-16707, https://doi.org/10.1364/OE.19.016697.

[9] T. Watanabe, Y. Kokubun, Ultra-large number of transmission channels in space division multiplexing using few-mode multi-core fiber with optimized air-hole-assisted double-cladding structure, Opt. Express 22 (2014) 8309-8319, https://doi.org/10.1364/OE.22.008309.

[10] Y. Choi, C. Yoon, M. Kim, T. Yang, C. Fang-Yen, R. Dasari, K. Lee, W. Choi, Scanner-free and wide-field endoscopic imaging by using a single multimode optical fiber, Phys. Rev. Lett. 109 (2012) 203901, https://doi.org/10.1103/PhysRevLett.109.203901.

[11] D. J. Richardson, J. Nilsson, W.A. Clarkson, High power fiber lasers: current status and future perspectives, J. Opt. Soc. Am. B 27 (2010) B63-B92, https://doi.org/10.1364/JOSAB.27.000B63.

[12] Z. Wang, D. Wang, F. Yang, L. Li, C. Zhao, B. Xu, S. Jin, S. Cao, Z. Fang, Er-doped mode-locked fiber laser with a hybrid structure of a step-index–graded-index multimode fiber as the saturable absorber, J. Lightwave Technol. 35 (2017) 5280-5285, https://doi.org/10.1109/JLT.2017.2768663.

[13] W. H. Renninger, F. W. Wise, Optical solitons in graded-index multimode fibres, Nat. Commun. 4 (2013) 1719, https://doi.org/10.1038/ncomms2739.

[14] L. Wright, D. Christodoulides, F. Wise, Controllable spatiotemporal nonlinear effects in multimode fibres, Nat. Photonics 9 (2015) 306-310, https://doi.org/10.1038/nphoton.2015.61.

[15] L. Wright, W. Renninger, D. Christodoulides, F. Wise, Spatiotemporal dynamics of multimode optical solitons, Opt. Express 23 (2015) 3492-3506, https://doi.org/10.1364/OE.23.003492.

[16] K. Krupa, A. Tonello, A. Barthélémy, V. Couderc, B. Shalaby, A. Bendahmane, G. Millot, S. Wabnitz, Observation of geometric parametric instability induced by the periodic spatial self-imaging of multimode waves, Phys. Rev. Lett. 116 (2016) 183901-183906, https://doi.org/10.1103/PhysRevLett.116.183901.

[17] L. G. Wright, S. Wabnitz, D. N. Christodoulides, F. W. Wise, Ultrabroadband dispersive radiation by spatiotemporal oscillation of multimode waves, Phys. Rev. Lett. 115 (2015) 223902-223907, https://doi.org/10.1103/PhysRevLett.115.223902.

[18] A. Eftekhar, Z. S. Eznaveh, H. Lopez-Aviles, S. Benis, J. Antonio-Lopez, M. Kolesik, F. Wise, R. Amezcua-Correa, D. Christodoulides, Accelerated nonlinear interactions in graded-index multimode fibers, Nat. Commun. 10 (2019) 1638, https://doi.org/10.1038/s41467-019-09687-9.

[19] Z. Liu, L. Wright, D. Christodoulides, F. Wise, Kerr self-cleaning of femtosecond-pulsed beams in graded-index multimode fiber, Opt. Lett. 41 (2016) 3675-3678, https://doi.org/10.1364/OL.41.003675.

[20] G. Lopez-Galmiche, Z. S. Eznaveh, A. Eftekhar, J. Lopez, L. Wright, F. Wise, D. Christodoulides, R. Correa, Visible supercontinuum generation in a graded index multimode fiber pumped at 1064 nm, Opt. Lett. 41 (2016) 2553-2556, https://doi.org/10.1364/OL.41.002553.

[21] L. Wright, Z. Liu, D. A. Nolan, M.-J. Li, D. Christodoulides, F. Wise, Self-organized instability in graded-index multimode fibre, Nat. Photonics 10 (2016) 771-776, https://doi.org/10.1038/nphoton.2016.227.

[22] D. Ceoldo, K. Krupa, A. Tonello, V. Couderc, D. Modotto, U. Minoni, G. Millot, S. Wabnitz, Second harmonic generation in multimode graded-index fibers: spatial beam cleaning and multiple harmonic sideband generation, Opt. Lett. 42 (2017) 971-974, https://doi.org/10.1364/OL.42.000971.

[23] R. Guenard, K. Krupa, R. Dupiol, M. Fabert, A. Bendahmane, V. Kermene, A. Desfarges-Berthelemot, J.-L. Auguste, A. Tonello, A. Barthélémy, G. Millot, S. Wabnitz, V. Couderc, Nonlinear beam self-cleaning in a coupled cavity composite laser based on multimode fiber, Opt. Express 25 (2017) 22219-22227, https://doi.org/10.1364/OE.25.022219.

[24] R. Guenard, K. Krupa, R. Dupiol, M. Fabert, A. Bendahmane, V. Kermene, A. Desfarges-Berthelemot, J.-L. Auguste, A. Tonello, A. Barthélémy, G. Millot, S. Wabnitz, V. Couderc, Kerr self-cleaning of pulsed beam in an Ytterbium doped

multimode fiber, Opt. Express 25 (2017) 4783-4792, https://doi.org/10.1364/OE.25.004783.

[25] K. Krupa, A. Tonello, B. M. Shalaby, M. Fabert, A. Barthélémy, G. Millot, S. Wabnitz, V. Couderc, Spatial beam self-cleaning in multimode fibres, Nat. Photonics 11 (2017) 237-241, https://doi.org/10.1038/nphoton.2017.32.

[26] R. Dupiol, K. Krupa, A. Tonello, M. Fabert, D. Modotto, S. Wabnitz, G. Millot, V. Couderc, Interplay of Kerr and Raman beam cleaning with a multimode microstructure fiber, Opt. Lett. 43 (2018) 587-590, https://doi.org/10.1364/OL.43.000587.

[27] K. Krupa, A. Tonello, V. Couderc, A. Barthélémy, G. Millot, D. Modotto, S. Wabnitz, Spatiotemporal light-beam compression from nonlinear mode coupling, Phys. Rev. A 97 (2018) 043836-043842, https://doi.org/10.1103/PhysRevA.97.043836.

[28] E. Deliancourt, M. Fabert, A. Tonello, K. Krupa, A. Desfarges-Berthelemot, V. Kermene, G. Millot, A. Barthélémy, S. Wabnitz, V. Couderc, Kerr beam self-cleaning on the $LP_{11}$ mode in graded-index multimode fibers, OSA Continuum 2 (2019) 1089-1096, https://doi.org/10.1364/OSAC.2.001089.

[29] E. Deliancourt, M. Fabert, A. Tonello, K. Krupa, A. Desfarges-Berthelemot, V. Kermene, G. Millot, A. Barthélémy, S. Wabnitz, V. Couderc, Wavefront shaping for optimized many-mode Kerr beam self-cleaning in graded-index multimode fiber, Opt. Express 27 (2019) 17311-17321, https://doi.org/10.1364/OE.27.017311.

[30] A. Fusaro, J. Garnier, K. Krupa, G. Millot, A. Picozzi, Dramatic acceleration of wave condensation mediated by disorder in multimode fibers, Phys. Rev. Lett. 122 (2019) 123902-123909, https://doi.org/10.1103/PhysRevLett.122.123902.

[31] K. Krupa, G. Garmendia Castañeda, A. Tonello, A. Niang, D. Kharenko, M. Fabert, V. Couderc, G. Millot, U. Minoni, D. Modotto, S. Wabnitz, Nonlinear polarization dynamics of Kerr beam self-cleaning in a graded-index multimode optical fiber, Opt. Lett. 44 (2019) 171-174, https://doi.org/10.1364/OL.44.000171.

[32] A. Niang, T. Mansuryan, K. Krupa, A. Tonello, M. Fabert, P. Leproux, D. Modotto, O. Egorova, A. Levchenko, D. Lipatov, S. Semjonov, G. Millot, V. Couderc, S. Wabnitz, Spatial beam self-cleaning and supercontinuum generation with Yb-doped multimode graded-index fiber taper based on accelerating self-imaging and dissipative landscape, Opt. Express 27 (2019) 24018-24028, https://doi.org/10.1364/OE.27.024018.

[33] E. V. Podivilov, D. Kharenko, V. Gonta, K. Krupa, O. Sidelnikov, S. Turitsyn, M. P. Fedoruk, S. Babin, S. Wabnitz, Hydrodynamic 2D turbulence and spatial beam condensation in multimode optical fibers, Phys. Rev. Lett. 122 (2019) 103902-103908, https://doi.org/10.1103/PhysRevLett.122.103902.

[34] A. Niang, A. Levchenko, S. Semjonov, D. Lipatov, S. Babin, V. Couderc, S. Wabnitz, D. Modotto, A. Tonello, F. Mangini, U. Minoni, M. Zitelli, M. Fabert, M. Jima, O. Egorova, Spatial beam self-cleaning in tapered Yb-doped GRIN multimode

fiber with decelerating nonlinearity, IEEE Photonics J. 12 (2020) 1-8, https://doi.org/10.1109/JPHOT.2020.2979938.

[35] B. Zhang, S. Ma, Q. He, J. Guo, Z. Jiao, B. Wang, Investigation on saturable absorbers based on nonlinear Kerr beam cleanup effect, Opt. Express 28 (2020) 6367-6377, https://doi.org/10.1364/OE.384376.

[36] K. Krupa, R. Fona, A. Tonello, A. Labruyère, B. M. Shalaby, S. Wabnitz, F. Baronio, A. B. Aceves, G. Millot, V. Couderc, Spatial beam self-cleaning in second-harmonic generation, Sci. Rep. 10 (2020) 7204, https://doi.org/10.1038/s41598-020-64080-7.

[37] S. Wabnitz, Y. Leventoux, A. Parriaux, O. Sidelnikov, G. Granger, L. Lavoute, M. Jossent, D. A. Gaponov, M. Fabert, A. Tonello, K. Krupa, A. Desfarges-Berthelemot, V. Kermene, G. Millot, S. Février, V. Couderc, Highly efficient few-mode spatial beam self-cleaning at 1.5μm, Opt. Express 28 (2020) 14333-14344, https://doi.org/10.1364/OE.392081.

[38] J. Lægsgaard, Spatial beam cleanup by pure Kerr processes in multimode fibers, Opt. Lett. 43 (2018) 2700-2703, https://doi.org/10.1364/OL.43.002700.

[39] K. Krupa, V. Couderc, A. Tonello, D. Modotto, A. Barthélémy, G. Millot, S. Wabnitz, Refractive index frofile tailoring of multimode optical fibers for the spatial and spectral shaping of parametric sidebands, J. Opt. Soc. Am. B-Opt. Phys. 36 (2019) 1117-1126, https://doi.org/10.1364/JOSAB.36.001117.

[40] K. Chiang, Stimulated Raman scattering in a multimode optical fiber: evolution of modes in Stokes waves, Opt. Lett. 17 (1992) 352-354, https://doi.org/10.1364/OL.17.000352.

[41] A. L. Gaeta, Catastrophic collapse of ultrashort pulses, Phys. Rev. Lett. 84 (2000) 3582-3585, https://doi.org/10.1103/PhysRevLett.84.3582.

[42] L. Tong, R. Gattass, J. Ashcom, S. He, J. Lou, M. Shen, I. Maxwell, E. Mazur, Subwavelength-diameter silica wires for low-loss optical wave guiding, Nature 426 (2004), 816-819, https://doi.org/10.1038/nature02193.

[43] M. Zitelli, F. Mangini, M. Ferraro, A. Niang, D. Kharenko, S. Wabnitz, High-energy soliton fission dynamics in multimode GRIN fiber, Opt. Express 28 (2020) 20473-20488, https://doi.org/10.1364/OE.394896.